\documentclass[aps,prl,twocolumn,showpacs,preprintnumbers]{revtex4-1} 
\pdfoutput=1

\usepackage{psfrag,graphicx}
\usepackage{dcolumn}
\usepackage{amsmath,amssymb,amsfonts}
\usepackage{bm}

\usepackage{hyperref}
\usepackage{braket}

\usepackage{subcaption} 

\newcommand{\SOI}{\textsc{soi}}
\newcommand{\pauli}[1]{\ensuremath{\sigma_\text{#1}}}
\newcommand{\hadamard}{\mathrm{H}}

\newcommand{\cnotvone}{\textsc{cnot}\textsubscript{v1}}
\newcommand{\cnotvtwo}{\textsc{cnot}\textsubscript{v2}}
\newcommand{\cnotloss}{\textsc{cnot}\textsubscript{L-DiV}}
\newcommand{\scenarioone}{scenario \textsc{i}}
\newcommand{\scenariotwo}{scenario \textsc{ii}}
\newcommand{\cnot}{\textsc{cnot}}
\newcommand{\swap}{\textsc{swap}}

\newcommand{\twoDEG}{\textsc{2deg}}
\newcommand{\on}[2]{\ensuremath{#1^{(#2)}}}
\newcommand{\scode}[1]{\ensuremath{\mathcal{S}_{#1}}}

\DeclareMathOperator{\Tr}{Tr}
\renewcommand\bra[1]{{\langle{#1}|}}
\makeatletter
\renewcommand\ket[1]{%
    \@ifnextchar\bra{\k@t{#1}\!}{\k@t{#1}}%
}
\newcommand\k@t[1]{{|{#1}\rangle}}
\makeatother

\begin{document}
\title{Benchmarks for approximate CNOTs based on a 17-qubit surface code}
\author{Andreas Peter}
\email{andreas.peter@strath.ac.uk}
\author{Daniel Loss}
\author{James R. Wootton}
\affiliation{Department of Physics, University of Basel, Klingelbergstrasse 82, CH-4056 Basel, Switzerland}

\begin{abstract}
    Scalable and fault-tolerant quantum computation will require error correction.
This will demand constant measurement of many-qubit observables, implemented using a vast number of \cnot{} gates. Indeed, practically all operations performed by a fault-tolerant device will be these \cnot{}s, or equivalent two-qubit controlled operations.
    It is therefore important to devise benchmarks for these gates that explicitly quantify their effectiveness at this task.
    Here we develop such benchmarks, and demonstrate their use by applying them to a range of differently 
    implemented controlled gates and a particular quantum error correcting code.
    Specifically, we consider spin qubits confined to quantum dots that are coupled either directly or
    via floating gates to implement the minimal 17-qubit instance of the surface code.
    Our results show that small differences in the gate fidelity can lead to large
    differences in the performance of the surface code.
    This shows that gate fidelity is not, in general, a good predictor of code performance.
\end{abstract}

\maketitle

\section{Introduction}

Two-qubit controlled operations are the workhorses of quantum algorithms.
In combination with single qubit rotations they can be used to implement a complete gate set.
The most widely considered controlled operations are the controlled-\textsc{not} 
(or \cnot{}), and gates that are equivalent to it by local unitaries~\cite{nielsen_quantum_2000}.

In practice, the \cnot{} needs to be implemented using the physical effects provided by a particular system.
The interactions that can be directly implemented by a system are compiled into a sequence of gates, designed to create an effective \cnot{}.
These gate sequences will only approximately implement a \cnot{},
both due to \emph{practical} limitations of experimental control and 
\emph{fundamental} limitations of approximating a gate by a limited set of other gates.
Therefore it is important to assess how the choice of a gate sequence to approximate a \cnot{} influences
the performance of a quantum algorithm.

To determine the quality of a \cnot{}, the standard means is to calculate the \emph{gate fidelity}.
While this provides some insight into how good an approximation a given gate is,
there is no information on the effects of the imperfections.
This has led to the fidelity being recognized as a not entirely trustworthy
means to compare quantum gates~\cite{bravyi_correcting_2017}.

Since one of the main tasks of the \cnot{} gates is the implementation of quantum error correction,
its performance in this context is especially important.
Ideally we would assess the performance in error correcting codes that are large enough for practical applications,
implementing a set of universal gates that could be applied fault-tolerantly.
However, the system sizes that need to be achieved for such a setup are far beyond the abilities of current
numerical techniques or experimental setups.

We will therefore consider a minimal working example:
The 17-qubit surface code which can both detect and correct quantum errors~\cite{tomita_low-distance_2014}.
The effects of both coherent and incoherent noise can be simulated for this code using a tensor network decomposition.
The results provide a direct insight into the performance of an approximate \cnot{} in an error correcting
code.

\section{Implementing CNOTs with Spin Qubits}

In order to provide specific and realistic examples for which calculations can be made,
we will restrict ourselves to \cnot{}s implemented on qubits formed by electrons confined to quantum dots.
At the most basic level, a quantum dot is a structure that \emph{confines} electrons to volumes
whose lengths are comparable to the wavelengths of the electrons.

In this work we consider \emph{lateral} quantum dots~\cite{kouwenhoven_electron_1997},
where an electron gas is strongly confined in one dimension, becoming an
effectively two-dimensional electron gas (\twoDEG{}).
This can be achieved at the boundary of two semiconductors
with different band gaps and suitable doping~\cite{etienne_two-dimensional_1987}.

Placing metallic gates on the \twoDEG{} and applying a negative potential on these gates,
the electrons can further be confined within the \twoDEG{} to specific islands: the quantum dots.
The quantum dots can be loaded with single electron precision and the \emph{electron's spin}
can be used to encode a \emph{qubit}.

We consider two ways to perform \cnot{} gates on these qubits.

\subsection{Loss-DiVincenzo CNOT}\label{ss:lossdivi}

A straightforward way to couple the electrons, and therefore the qubits,
in two neighboring quantum dots is \emph{transient Heisenberg coupling}~\cite{loss_quantum_1998}.
By lowering the potential between the two quantum dots, the wavefunctions of the qubits
overlap which leads to a coupling of the form
\begin{equation}\label{eq:heisenbergcoupling}
H_\text{s}(t) = J(t) \vec{\sigma}^{(1)} \cdot \vec{\sigma}^{(2)},
\end{equation}
where $J(t)$ is a time-dependent exchange constant and $\vec{\sigma}$
are vectors with the Pauli operators as elements where the superscripts
show on which qubit they act.

Applying the gate for a duration $\tau_\text{s}$  such that 
${\int_0^{\tau_\text{s}} J(t) dt = \tfrac{\pi}{2}}$ results
in the application of $H_\text{s}$ being equivalent to a $\sqrt{\text{\swap{}}}$ gate,
i.e. a gate which, when applied twice, interchanges the state of the two qubits.

The $\sqrt{\text{\swap{}}}$ gate can be used to construct a \cnot{} using the sequence
\begin{equation}\label{eq:cnotloss}
\cnotloss{} =
    \sqrt{\pauli{z}}^{(1)} \sqrt{-\pauli{z}}^{(2)} \sqrt{\text{\swap{}}} \;
    \pauli{z}^{(1)} \sqrt{\text{\swap{}}} .
\end{equation}
The required single qubit rotations can be achieved in the quantum dot
by applying appropriate magnetic fields and
using electron spin resonance~\cite{kloeffel_prospects_2013}.

As written in \autoref{eq:cnotloss}, the \cnot{} is exact.
For a more realistic analysis, we replace the unitary $H_\text{s}$ in \autoref{eq:cnotloss}
with a \emph{quantum channel} that takes interactions with the environment during
the application of $H_\text{s}$ into account.

Using Markov and Born approximations, the channel
\begin{equation}\label{eq:v}
\mathcal{V}(t) = \exp[-(t-\tau_\text{s}) \mathcal{K}_3 ]  \,
    \mathcal{U}_\text{s}(\tau_\text{s}) (\openone - \mathcal{K}_2),
\end{equation}
is derived in \cite{loss_quantum_1998}.
In \autoref{eq:v}, $\tau_\text{s}$ is the time the Hamiltonian \autoref{eq:heisenbergcoupling}
is applied, $t > \tau_\text{s}$ is the total time considered and $\mathcal{U}_\text{s}$, $\mathcal{K}_3$, 
and $\mathcal{K}_2$ are the channels for the unitary application of $H_\text{s}$, effect of the environment
during the application of $H_\text{s}$ and the correction to the initial state due to the environment,
respectively.

Cast into the form of matrices, the channel in \autoref{eq:v} can be expressed as a function
of three parameters: $t$, $\Gamma$ and $\Delta$, with $\Gamma$ being the spin 
relaxation rate and $\Delta$ a phase shift. Both are functions of the
coupling strength between spin qubit and environment.
For the derivation and more details see \cite{loss_quantum_1998}.

This implementation of the \cnot{} requires short-range coupling,
and a technically challenging close packing of the qubits.

\subsection{Floating Gate CNOT}\label{ss:float}

These challenges can be avoided using other methods to implement the \cnot{},
such as the capacitive coupling of quantum dots via floating gates~\cite{trifunovic_long-distance_2012}.

This implementation takes advantage of electron spins coupling electrostatically,
since the electrons feel each others charges.
Specifically, the charge degree of freedom is coupled to the orbital motion,
and the orbital motion is coupled to the spin via the \emph{spin-orbit interaction}
(\SOI{})~\cite{trif_spin-spin_2007}.

The charges are screened by gates and the \twoDEG{} between quantum dots,
and so electrostatic coupling cannot be used directly for long-distance coupling.
Instead it can be mediated by floating gates:
two flat metallic disks connected by a thin metallic wire.
Placing connected disks near a quantum dot each,
the mirror charge induced in either disk by one 
electron charges the other disk with opposite sign and acts on
the other electron.
The resulting coupling decays only weakly with the wire distance, enabling long-distance coupling
and loosening the requirements on the quantum dot arrangements.

A complete analysis of the electrostatic forces, see \cite{trifunovic_long-distance_2012} for details,
yields an effective Hamiltonian of the form
\begin{equation}\label{eq:hamilfloatinggates}
    H = E_z(\pauli{z}^{(1)} + \pauli{z}^{(2)}) + J_{12} 
    (\vec{\sigma}^{(1)} \cdot \vec{\gamma}) (\vec{\sigma}^{(2)} \cdot \vec{\gamma}),
\end{equation}
where $J_{12}$ is a coupling constant that depends on both magnetic field and the details of the metallic gate and
\begin{equation}
\vec{\gamma} = (\alpha_\text{D} \cos ( 2 \gamma) , -
\alpha_\text{R} - \alpha_\text{D} \sin ( 2 \gamma ), 0) , 
\end{equation}
with $\alpha_\text{R}$ and $\alpha_\text{D}$ being the \emph{Rashba} and \emph{Dresselhaus} \SOI{}-strength, respectively, and
$\gamma$ is the angle between the crystallographic axis along the $[100]$ direction and the axis along the wire.

The Hamiltonian in \autoref{eq:hamilfloatinggates} can be approximated by
\begin{equation}
H' = \tfrac{J_{12}}{2} |\gamma_x|^2 \left( \pauli{x}^{(1)}
\pauli{x}^{(2)} +\pauli{y}^{(1)} \pauli{y}^{(2)}\right)
+E_z(\pauli{z}^{(1)} + \pauli{z}^{(2)}), 
\end{equation}
if $E_z \gg J_{12} |\gamma_x|^2$ and assuming that the magnetic field is perpendicular to the \twoDEG{}.

We can use $H'$ to implement a $\sqrt{ \pauli{x} \pauli{x}}$ gate in two ways, with two and four applications of $H'$, respectively:
\begin{align}
\sqrt{\pauli{x}\pauli{x}}_\text{v1} &= 
\exp[\i(\pauli{z}^{(1)} + \pauli{z}^{(2)}) E_z t]
\exp[-\i H' t] \pauli{x}^{(1)}\\\notag
&\qquad\exp[\i(\pauli{z}^{(1)} + \pauli{z}^{(2)}) E_z t]
\exp[-\i H' t] \pauli{x}^{(1)}\\
\sqrt{\pauli{x}\pauli{x}}_\text{v2} &= 
\pauli{x}^{(2)} \exp[-\i H' \tfrac{t}{2}] \pauli{x}^{(1)}
\pauli{x}^{(2)} \exp[-\i H' \tfrac{t}{2}]\\\notag
&\qquad \pauli{x}^{(2)}
\exp[-\i H' \tfrac{t}{2}] \pauli{x}^{(1)}
\pauli{x}^{(2)} \exp[-\i H' \tfrac{t}{2}]\, ,
\end{align}

where the application time $t$ is given by
\begin{equation}
\label{eq:fgtime}
    t= \frac{\pi}{4 J_{12} \left(\gamma_x^2 + \gamma_y^2\right)}.
\end{equation}

Given an implementation for $\sqrt{\pauli{x} \pauli{x}}$ it is straightforward to implement a \cnot{} as
\begin{equation}
\text{\cnot{}} = \on{\sqrt{\pauli{z}}}{1} \on{\sqrt{\pauli{x}}}{2}
\on{\hadamard{}}{1} \sqrt{\pauli{x} \pauli{x}}\, \on{\hadamard{}}{1},
\end{equation}

where $\hadamard{}$ is the \emph{Hadamard} gate.
        
This implementation is only exact for $H'$. For the real Hamiltonian of
\autoref{eq:hamilfloatinggates}, the gate, while unitary, will not 
exactly be a \cnot{} but instead an approximate channel \cnotvone{} or
\cnotvtwo{} for two and four Hamiltonian applications, respectively.
To define the nature of these approximate \cnot{}s it is only necessary to choose
the parameters of the floating gate Hamiltonian.

\section{17-qubit Surface Code}\label{subsec:s17}

\begin{figure}[htb]
    \includegraphics{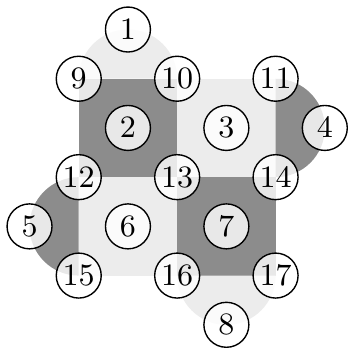}
    \caption{\label{fig:labelings17}
    A surface code defined on a $3 \times 3$ array of physical data qubits.
    The numbering used for the qubits in this work is shown.}
\end{figure}

To protect quantum information against physical noise induced by an environment,
\emph{quantum error correcting codes} can be used~\cite{lidar_quantum_2013}.
For our purposes we use nine \emph{physical data} qubits in whose Hilbert space
we encode one \emph{logical} qubit.
The nine data qubits are arranged on a $3\times3$ lattice as the
points labeled $9$ to $17$ in \autoref{fig:labelings17}.
These qubits span a $2^9$-dimensional Hilbert space.

The 17 qubit surface code, \scode{17}, is defined by the set of eight mutually
commuting \emph{stabilizer} operators
\allowdisplaybreaks[0]
\begin{multline}
    \label{eq:s17stab}
    \scode{17}=\{
        X_2=\on{\pauli{x}}{9} \on{\pauli{z}}{10}
            \on{\pauli{z}}{12} \on{\pauli{x}}{13},\\
        X_7=\on{\pauli{x}}{13} \on{\pauli{z}}{14}
            \on{\pauli{z}}{16} \on{\pauli{x}}{17},
        X_4=\on{\pauli{x}}{11} \on{\pauli{z}}{14},\\
        X_5=\on{\pauli{z}}{12} \on{\pauli{x}}{15},
        Z_3=\on{\pauli{x}}{10} \on{\pauli{z}}{11}
            \on{\pauli{z}}{13} \on{\pauli{x}}{14},\\
        Z_6=\on{\pauli{x}}{12} \on{\pauli{z}}{13}
            \on{\pauli{z}}{15} \on{\pauli{x}}{16},
        Z_1=\on{\pauli{z}}{9} \on{\pauli{x}}{10},\\
        Z_8=\on{\pauli{x}}{16} \on{\pauli{z}}{17}
        \},
\end{multline}
\allowdisplaybreaks
where the operators are grouped into $X$ and $Z$ operators.
We can restrict data qubits to be in the mutual $+1$ eigenstate
of all operators in \scode{17}. The resulting Hilbert space --- the \emph{code space} ---
is $2^{9-8}=2$-dimensional and can encode \emph{one} logical qubit.

We define an operator 
$Z_L = \on{\pauli{x}}{10} \on{\pauli{z}}{13} \on{\pauli{x}}{16}$
that commutes with all stabilizers of \scode{17}.
Identifying it with the logical $Z$-operator lets us define the logical $\ket{0}$
and $\ket{1}$ states as the $+1$ and $-1$ eigenstates of $Z_L$ in the code space, respectively.
The logical $X$-operator is $X_L = \on{\pauli{z}}{12} \on{\pauli{x}}{13} \on{\pauli{z}}{14}$
which commutes with all elements of \scode{17} and anticommutes with $Z_L$.
Since stabilizers do not change logical states, 
the logical operators are only unique up to multiplication with stabilizers.

\scode{17} is an example of a \emph{surface code} which belongs to the class
of \emph{stabilizer codes}. For detailed information on both topics
see references \cite{lidar_quantum_2013} and \cite{dennis_topological_2001},
but for our purpose it is enough to note that \scode{17} is the smallest useful
realization of a surface code considering both number of qubits and number of operations.

Having defined the logical state and the code space with \scode{17},
it is easy to see that any single-qubit error on one of the data qubits
will be projected to a Pauli operator upon measurement of a stabilizer and any
Pauli operator on one of the qubits anticommutes with at least one stabilizer.
Thus by measuring the stabilizers, single qubit errors can be \emph{detected}.

To enable measurement without genuine 4-qubit interactions, we use 8
\emph{ancilla} qubits --- one associated with each stabilizer.
The ancillas and their corresponding stabilizers are shown in \autoref{fig:labelings17}
as the qubits $1$ to $8$ and the plaquettes they are on, respectively.
The ancillas are entangled to each qubit their respective stabilizers act
on non-trivially such that their measurement is equivalent to the measurement
of the stabilizer.

\begin{figure}[htb]
    \includegraphics{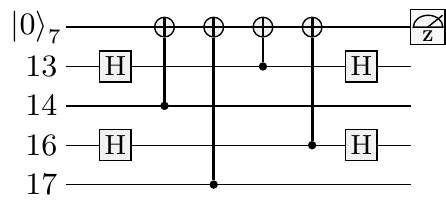}
    \caption{Measurement of $X_7 \in \text{\scode{17}}$ via ancilla $7$.}
    \label{fig:measureanc}
\end{figure}

As an example, consider the indirect measurement of $X_7$ shown
in \autoref{fig:measureanc} via an entangled ancilla qubit requiring
only two-qubit gates --- \cnot{}s --- and one-qubit measurement.
The sequence is chosen such that the concurrent measurement of all stabilizers
with the same sequence does not move errors around but note
that with concurrent measurements, Hadamards would need to be applied
directly before and after the \cnot{}s.

\subsection{Decoding}\label{subsec:Decoding} 

\begin{figure}[htb]
    \captionsetup[subfigure]{aboveskip=-8pt}
        \begin{subfigure}{.1 \textwidth}
            \centering
            \includegraphics{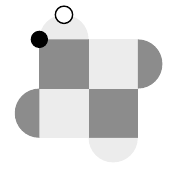}
            \vspace*{-1mm}
            \caption{{}}
        \end{subfigure}
        ~
        \begin{subfigure}{.1 \textwidth}
            \centering
            \includegraphics{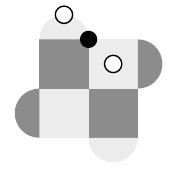}
            \vspace*{-1mm}
            \caption{{}}
        \end{subfigure}
        ~
        \begin{subfigure}{.1 \textwidth}
            \centering
            \includegraphics{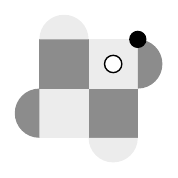}
            \vspace*{-1mm}
            \caption{{}}
            \label{fig:s17decoding11}
        \end{subfigure}
        ~
        \begin{subfigure}{.1 \textwidth}
            \centering
            \includegraphics{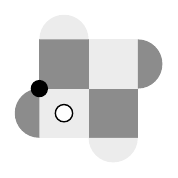}
            \vspace*{-1mm}
            \caption{{}}
            \label{fig:s17decoding21}
        \end{subfigure}

        \begin{subfigure}{.1 \textwidth}
            \centering
            \includegraphics{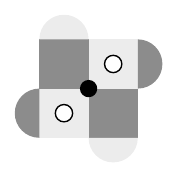}
            \vspace*{-1mm}
            \caption{{}}
        \end{subfigure}
        ~
        \begin{subfigure}{.1 \textwidth}
            \centering
            \includegraphics{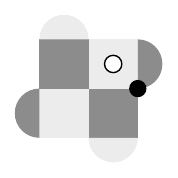}
            \vspace*{-1mm}
            \caption{{}}
            \label{fig:s17decoding12}
        \end{subfigure}
        ~
        \begin{subfigure}{.1 \textwidth}
            \centering
            \includegraphics{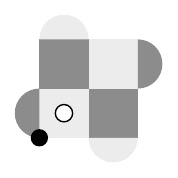}
            \vspace*{-1mm}
            \caption{{}}
            \label{fig:s17decoding22}
        \end{subfigure}
        ~
        \begin{subfigure}{.1 \textwidth}
            \centering
            \includegraphics{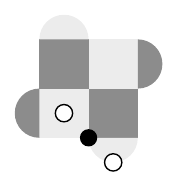}
            \vspace*{-1mm}
            \caption{{}}
        \end{subfigure}
         
        \begin{subfigure}{.1\textwidth}
            \centering
            \includegraphics{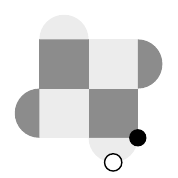}
            \vspace*{-1mm}
        \caption{{}}
        \end{subfigure}
    \caption{\label{fig:s17decoding}
    The syndrome for all one-qubit errors that \scode{17} can detect by $Z$-stabilizer measurement.
    Errors are shown as black dots, $-1$ syndromes as white dots with black boundaries.
    The graphic for $X$-stabilizers is the same with syndrome and errors rotated by $90$ degrees. }
\end{figure}

From the previous section it is clear that \scode{17} can \emph{detect} any 
one-qubit error on one of the 9 data qubits.
The measurement of the stabilizers gives us 8 bits of information which, adding
the measurement result of a logical operator, we bundle together as the \emph{syndrome}.
Taking a syndrome and extracting the most likely logical state is the process of \emph{decoding},
which enables us to \emph{correct} errors.

In Fig. \ref{fig:s17decoding} we can see the syndrome resulting from each one-qubit
error detectable by $Z$-stabilizers. The error, which is projected to a Pauli
operator, is detected by the part of the stabilizer which anticommutes with its effect.
While some errors lead to a unique syndrome and can be undone straight-forwardly,
others, such as \ref{fig:s17decoding11} and \ref{fig:s17decoding12}, cause the same syndrome.
It is, however, easy to see that it does not matter which one of the possible errors are undone
since the corrections of both errors are equivalent up to stabilizers.

For more than one Pauli error, decoding might fail to reveal the correct logical state.
This is clear since logical operators themselves can be mimicked by three Pauli errors,
yet are invisible to stabilizer measurements.

The above description relies on the measurements being perfect which implies both perfect
entangling gates --- \cnot{}s --- and noiseless ancillas.
In a physical implementation these assumptions are clearly not realistic.

To enable error correction under these conditions,
multiple measurement rounds in quick succession are necessary
to construct a three-dimensional syndrome which contains information about both
data and ancilla qubit errors  --- the latter showing themselves as successive
flips of the syndrome.

The final syndrome measurement round is done at the time of readout,
using the results from direct measurement of the data qubits.
Any imperfections in the measurement for this can be modeled 
as additional noise preceding a perfect measurement.
This final, effectively perfect, syndrome measurement round ensures 
that it is possible that single qubit measurement noise is detected and corrected.

For large codes --- in time, space or both --- the decoding requires sophisticated algorithms~\cite{fowler_minimum_2015}.
However, the system considered is compact enough that a simple and optimal procedure based on lookup-tables is possible~\cite{wootton_proposal_2017}.

The lookup-table results from many runs of a surface code for multiple possible encoded states,
obtained either experimentally or from simulations.
The final result $\bm{s}$, which is comprised of the syndrome and the result of a final logical measurement,
is recorded for each encoded state $\ket{\phi}$.
This data is then used to calculate the conditional probabilities,
$p_{\ket{\phi}}(\bm{s})$, which form the lookup-table.

If we are interested to measure in the computational basis,
we take $\ket{\phi} = \ket{0}$ and $\ket{\phi} = \ket{1}$.
A state can then be corrected by taking the syndrome and choosing whichever 
logical state is more likely given that syndrome according to the conditional
probabilities calculated beforehand.

The associated computational basis state of the probability chosen is then the \emph{best possible guess}
as to which state the system should be in without the effect of noise, e.g.  after a computation.
Correction can then be applied to match the measured logical state with
the probable logical state~\cite{darmawan_tensor-network_2017}.

The lookup-table not only provides us the best guess for the \emph{true} state
of a system but also its probability and thus the probability of that guess being wrong.

Note that the \scode{17} code is able to detect and correct all forms of single qubit error
that can occur during its implementation, whether they be from interaction with the environment,
imperfection in measurements or faulty implementation of gates.
However, it is the last of these that we will primarily focus on in this work.
Our main benchmarks will be measures of how well a code performs when the main
noise present is that associated with the implementation of a \cnot{}.

\section{Calculations}

\subsection{Fidelity}

To assess the quality of an approximate \cnot{} we can use the \emph{gate fidelity} of
approximate and exact \cnot{} defined as 
\begin{equation}
F(\mathcal{E},\mathcal{C}) = \min_{\ket{\psi}} F\big(
\mathcal{E}(\ket{\psi} \bra{\psi}),
\mathcal{C}(\ket{\psi} \bra{\psi})\big),
\end{equation}
where $\mathcal{E}$ and $\mathcal{C}$ are the channels for the approximate and exact
\cnot{}, respectively, and the minimum is taken over all pure states $\ket{\psi}$.
The fidelity $F$ for two states $\rho$ and $\sigma$ is defined as
\begin{equation}\label{eq:fidelity}
F(\rho,\sigma) = \Tr \left( \sqrt{\sqrt{\rho} \sigma \sqrt{\rho}} \right).
\end{equation}
$F$ is symmetric in its inputs, invariant under unitary transformations, and 
obeys ${ 0 \leq F(\rho,\sigma) \leq 1}$, where equality with $0$ implies
orthogonal support while equality with $1$ implies equality of states.
Additionally, \autoref{eq:fidelity} obeys \emph{strong concavity} which implies 
that the minimization need only be performed over pure states.

To limit computational cost we restrict the calculation further and only consider
the tensor products of eigenvectors of the Pauli operators.
Thus our result is strictly speaking only an upper bound on the gate fidelity but
more elaborate searches showed little difference with our calculated fidelities.

\subsection{Performance in a Surface Code}

To assess the performance of an approximate \cnot{} in a surface code,
we came up with two scenarios which we can handle numerically and deem
realistic for experimental implementation in the near future.

For both scenarios we choose the probability of failed decoding in \scode{17} as the
measure of its usefulness which can be calculated using a lookup-table as explained above.
To arrive at the lookup-table we calculate the probabilities of all possible syndromes
conditional on the initial logical state.

From the lookup-table it is straightforward to extract the probability of a logical error
by summing over the probabilities of all wrong decodings.
The result is the probability of wrong decoding or failed error correction.

For noise we consider, test calculations showed that there was not a significant
difference between results for the initial logical state being in either of the
computational basis states.
We therefore consider only initial logical state $\ket{0}$.

\subsection{Noise Model}

In the context of quantum computation,
any effects from the environment outside our direct control must be interpreted as \emph{noise};
the environment changes our carefully crafted quantum states in unwanted ways.

Simulation of noise would ideally be done by determining the full
time evolution of both system and environment.
The environment could then be traced out to obtain a final noise channel $\mathcal{E}$
which maps the state at an initial time $t_0$ to that at the later time $t$,
\begin{align} \nonumber
\rho_\text{s}(t) &= \mathcal{E}(\rho_{\text{s}}(t_0)).
\end{align}
Such a full simulation is not feasible for realistic noise,
and so determining $\mathcal{E}$ usually requires strong assumptions about the nature of the environment and
its entanglement with the system.

We will consider the representation of noise channels in terms of Kraus operators.
We therefore consider representations in terms of operators $K_j$ such that
\begin{align} \nonumber
\rho_\text{s}(t) &= \mathcal{E}(\rho_{\text{s}}(t_0)) \\
&= \sum_j K_j \,\, \rho_{\text{s}}(t_0) \,\, K_j^\dagger.
\end{align}

For our calculations we consider bit and phase flip and depolarizing noise.
While the former are primarily toy models that are easy to parametrize,
the latter provides a more realistic picture with parameters that
depend on physical considerations.

In terms of Kraus operators, bit and phase flip are represented as
\begin{equation}\label{eq:bitflip}
    K_1^{\text{bit}} = \sqrt{1-p}\, \openone ,\qquad K_2^{\text{bit}} = \sqrt{p}\, \pauli{x}
\end{equation}
and
\begin{equation}\label{eq:phaseflip}
K_1^{\text{phase}} = \sqrt{1-p}\, \openone , \qquad K_2^{\text{phase}}  = \sqrt{p}\, \pauli{z} ,
\end{equation}
respectively.
As expressed in \autoref{eq:bitflip} and \autoref{eq:phaseflip}, the two errors flip the 
state or the phase with a probability $p$ and are related by a simple base-change via a
Hadamard.

Depolarization describes a type of error that erases \emph{all} information about 
a state with a probability $p$, leaving us with the completely mixed state,
or leave the state undisturbed.
It can be represented by a channel
\begin{equation}\label{eq:depolgen}
    \mathcal{E}(\rho) = (1-p_\text{depol}) \rho + \tfrac{p_\text{depol}}{2} \openone.
\end{equation}
\autoref{eq:depolgen} can be expressed in terms of Kraus operators as
\begin{equation}\label{eq:depol}
    K_1 = \sqrt{1-\tfrac{3}{4} p_\text{depol}}\, \openone,  \qquad K_{i} = \sqrt{\tfrac{p_\text{depol}}{4}}\, \pauli{i}, 
\end{equation}
where $i$ goes from $2$ to $4$ and over the Pauli operators, respectively.

The channels used for the \cnot{}s will also be brought into the form of Kraus operators.
The effects of these are then simulated using tensor network techniques \cite{darmawan_tensor-network_2017}.

Our main aim in this study is to investigate how well the gate fidelity captures the
performance of a \cnot{} in an actual surface code.
Parameters for \cnot{}s have therefore been chosen such that the \emph{gate fidelity}
of the \cnot{}s have a nontrivial dependence on the parameters.

For the \emph{floating gate} \cnot{}s, we choose as parameters
the ratio between Zeeman energy and coupling constant
$R=E_z /\left(J_{12} \gamma_x^2\right)$ and $\gamma=\gamma_y/\gamma_x$.
We look at $R \in (30,35)$ and $\gamma \in (0,1)$ since for
these parameters the gate fidelity of  both \cnotvone{} and
\cnotvtwo{} shows interesting dependence on the parameters.

For the Loss-DiVincenzo \cnot{} we use as parameters
the time since the beginning of the interaction $t>1$ in units
of the interaction time $\tau_\text{s}$ and the decoherence
parameter $\Gamma$.
We look at $t\in (1,1.1)$ and $\Gamma \in (0.007,0.027)$ to
stick close to the values used in \cite{loss_quantum_1998} and keep
$\Delta$ fixed at $-0.0145$.

\subsection{Scenarios}

\begin{figure}[hbtp]
\begin{subfigure}{0.45 \columnwidth}
    \hspace*{-0.8cm}
    \includegraphics{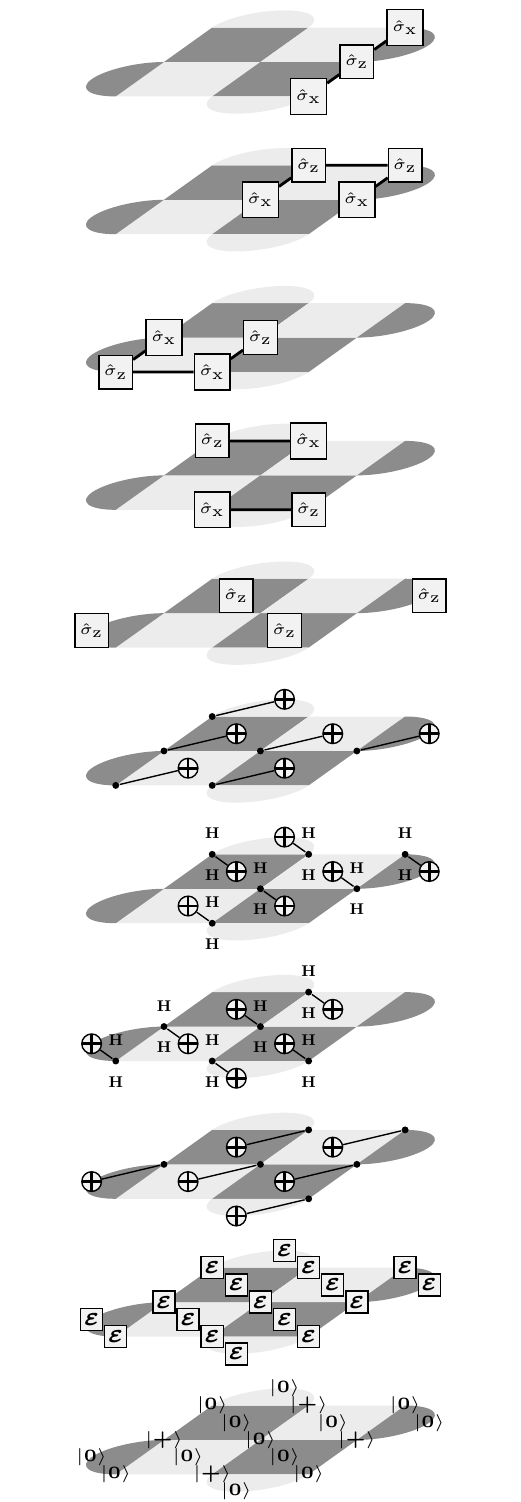}
    \caption{\scenarioone{}}
    \label{fig:algscenario1}
\end{subfigure}
\begin{subfigure}{0.45 \columnwidth}
    \hspace*{-0.8cm}
    \includegraphics{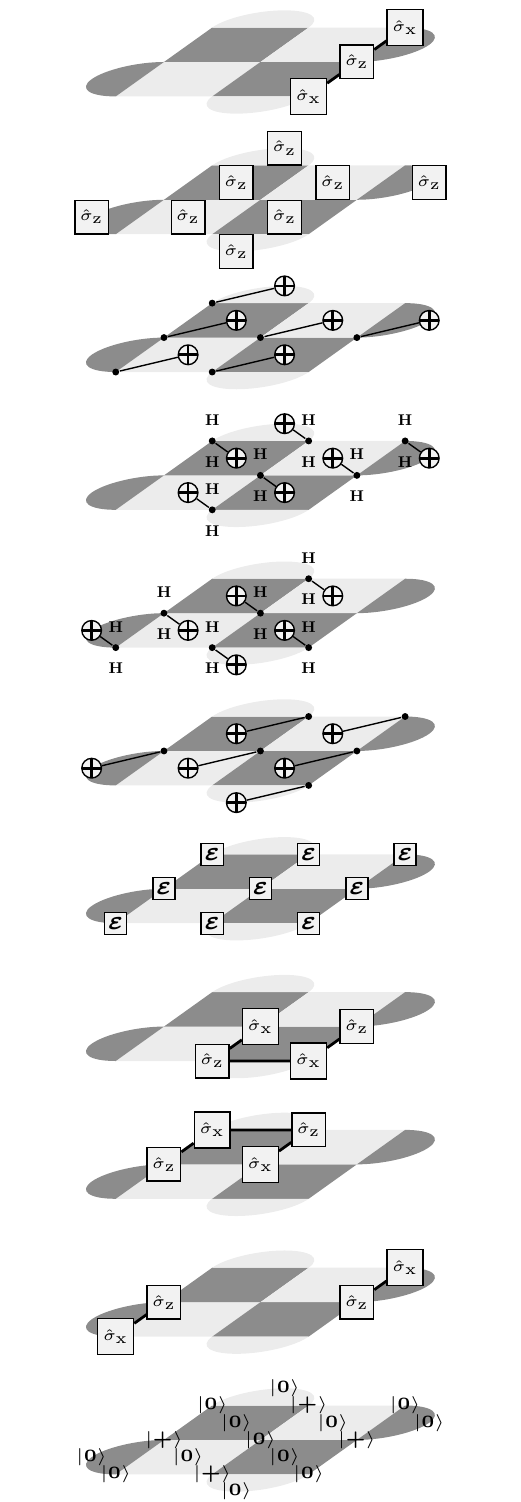}
    \caption{\scenariotwo{}}
    \label{fig:algscenario2}
\end{subfigure}
\caption{A depiction of the steps of the two scenarios used to assess the effectiveness of \cnot{}s.
The boxes with Paulis are projectors such that the projectors after (above) the \cnot{}s correspond to measurements with
the outcome determined by the projections. The corresponding probability of the measurement can then be extracted
by the norm of the state.}
\label{fig:algscenarios}
\end{figure}

\subsubsection{Scenario I}

The surface code \scode{17} is initialized as an eigenstate of
the $Z$ stabilizers (see \autoref{eq:s17stab}),
which can be achieved by starting in a product state of $\ket{0}$ and subsequently
applying Hadamards on qubits $10$, $12$, $14$, and $16$.
An initial round of noise is applied to \emph{all} qubits as \emph{depolarizing noise}
with strength $p_\text{depol} = p_\text{init}$ to mimic the preparation noise of the experiment.

Using the \cnot{} chosen, one round of syndrome measurement is applied where
Hadamards are applied to data qubits before and after measurement if
they are to be measured in the $\pauli{x}$ basis.
The ancillas of the $Z$-stabilizers are measured and for a final
readout each data qubit is measured in the basis that commutes with the $Z$-stabilizers acting on it.

The results of these measurements are then used to deduce the values of the $Z$-stabilizers,
effectively forming a final round of syndrome measurement, as well as the final value of the $Z$ logical operator.
All other information is discarded.

The results are therefore made up of eight bits of syndrome information from two rounds, and one bit of logical information.
From this syndrome we can calculate the conditional probabilities we need to get the code performance for \scenarioone{}.

This is the same setup we considered in \cite{wootton_proposal_2017}, where we considered 
exact \cnot{}s affected by different single-qubit noise channels.

\subsubsection{Scenario II}

The first scenario gives us information about just one type of stabilizers,
measured once indirectly and once directly. 
To extract information about the effect of using both sets of stabilizers and
within the constraints of our computational resources, \scenariotwo{} 
includes a full measurement of all stabilizers via the ancillas but
no final direct stabilizer measurement as in \scenarioone{}.
\scode{17} is initialized as a logical state, i.e. as a mutual $+1$ eigenstate of all operators
in \scode{17}.
This is achieved starting in the same eigenstate of all $Z$ operators as in \scenarioone{}
and subsequent application of projectors into $X$-stabilizer eigenstates.

Preparation noise is applied to the data qubits as 
depolarizing noise as before with strength $p_\text{init}$.
After that we apply one round of syndrome measurement using the chosen \cnot{}s.
This round ends with the measurement of \emph{all} ancillas to get an eight bit syndrome.

The final readout consists simply of the logical $Z$ measurement.
All other information is discarded.
The results are therefore again made up of eight bits of syndrome information,
and one bit of logical information.

Since we only measure one round of ancillas, we can not correct ancilla errors.
This is the reason preparation noise is only applied to the data qubits.
All ancilla errors will then be due to the imperfect channels that act as noisy \cnot{}s.

\section{Results}

Tensor network techniques were used to calculate the probabilities
of each possible syndrome in the two scenarios, inspired by the
approach in \cite{darmawan_tensor-network_2017}.
From these simulations, the probability of a logical error, $p_\text{code}$,
was calculated for both scenarios and all \cnot{}s described above.
This allowed us to determine how $p_\text{code}$ depends
on the imperfections of the \cnot{} gates used, and how this relates to the fidelity.

\subsection{Results for Scenario I}

We will now look at the code performance as a function of the gate infidelities, i.e. $(1 - \text{fidelity})$
of the \cnot{}s, focusing first on \scenarioone{}.
To do this we construct \cnot{}s using multiple different coupling parameters.
For each we calculate both $p_\text{code}$ and the gate infidelities of the \cnot{}s,
and plot these against each other to determine the dependence of the code performance on gate infidelity.

\begin{figure}[htbp]
    \begin{subfigure}{0.32 \columnwidth}
        \includegraphics{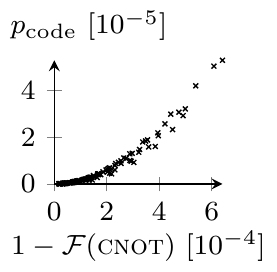}
        \includegraphics{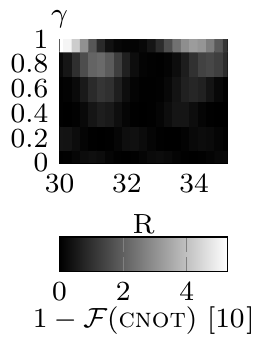}
        \caption{$p_\text{init}=0$}
    \end{subfigure}
    \begin{subfigure}{0.32 \columnwidth}
        \includegraphics{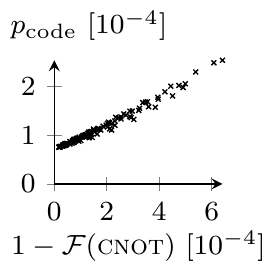}
        \includegraphics{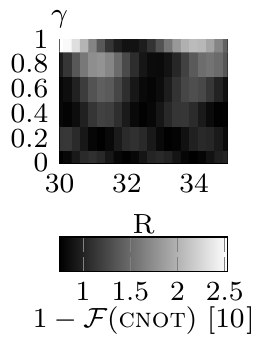}
        \caption{$p_\text{init}=0.002$}
    \end{subfigure}
    \begin{subfigure}{0.32 \columnwidth}
        \includegraphics{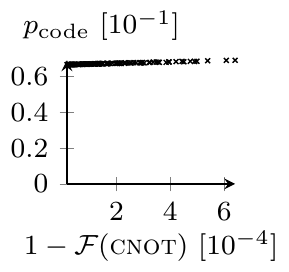}
        \includegraphics{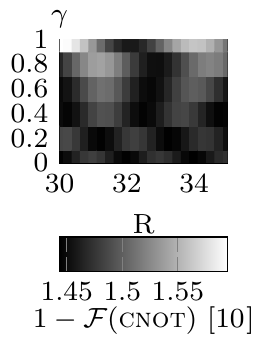}
        \caption{$p_\text{init}=0.07$}
    \end{subfigure}
    \caption{\label{fig:v1_s1}
    Plots of $p_\text{code}$ for \scenarioone{} versus the infidelity for \cnotvone{}.
    Different curves correspond to different values of $p_\text{init}$.}
\end{figure}

\begin{figure}[htbp]
    \begin{subfigure}{0.32 \columnwidth}
        \includegraphics{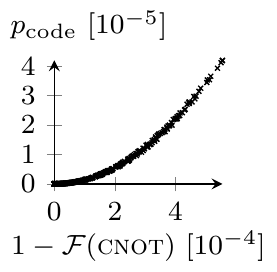}
        \includegraphics{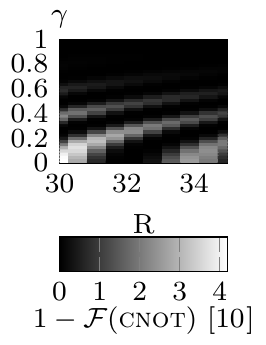}
        \caption{$p_\text{init}=0$}
    \end{subfigure}
    \begin{subfigure}{0.32 \columnwidth}
        \includegraphics{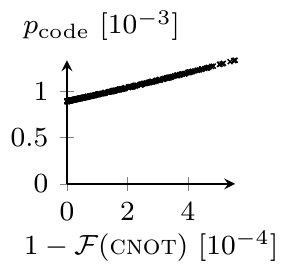}
        \includegraphics{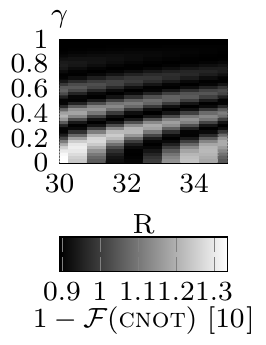}
        \caption{$p_\text{init}=0.007$}
    \end{subfigure}
    \begin{subfigure}{0.32 \columnwidth}
        \includegraphics{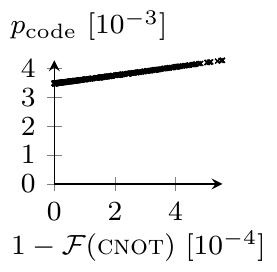}
        \includegraphics{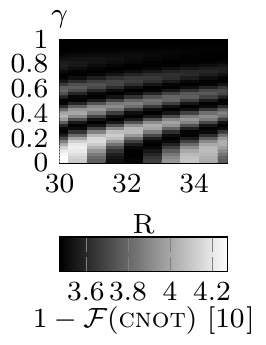}
        \caption{$p_\text{init}=0.014$}
    \end{subfigure}
    \caption{\label{fig:v2_s1}
    Plots of $p_\text{code}$ for \scenarioone{} versus the infidelity for \cnotvtwo{}.
    Different curves correspond to different values of $p_\text{init}$.}
\end{figure}

\begin{figure}[htbp]
    \begin{subfigure}{0.32 \columnwidth}
        \includegraphics{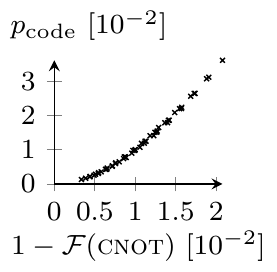}
        \includegraphics{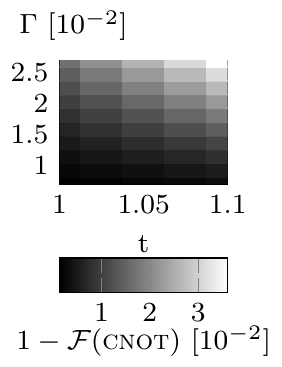}
        \caption{$p_\text{init}=0$}
    \end{subfigure}
    \begin{subfigure}{0.32 \columnwidth}
        \includegraphics{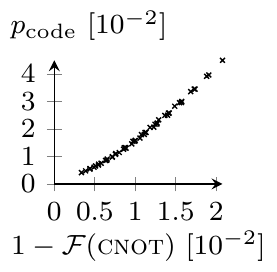}
        \includegraphics{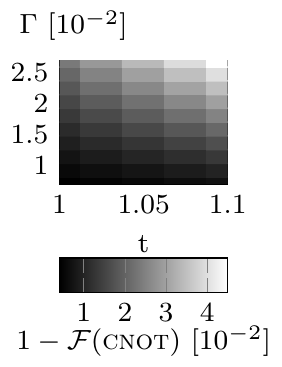}
        \caption{$p_\text{init}=0.003$}
    \end{subfigure}
    \begin{subfigure}{0.32 \columnwidth}
        \includegraphics{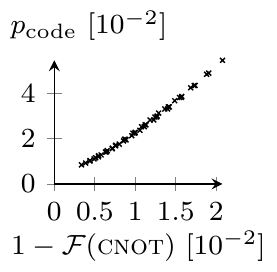}
        \includegraphics{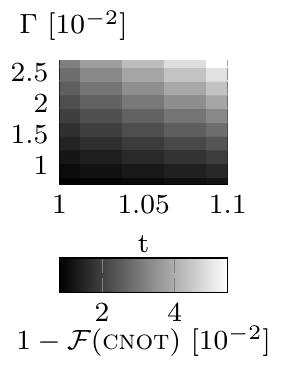}
        \caption{$p_\text{init}=0.007$}
    \end{subfigure}
    \caption{\label{fig:loss_s1}
    Plots of $p_\text{code}$ for \scenarioone{} versus the infidelity for \cnotloss{}.
    Different curves correspond to different values of $p_\text{init}$.}
\end{figure}

The dependence of $p_\text{code}$ on the infidelity of a \cnot{} is shown in  Fig. \ref{fig:v1_s1}, Fig. \ref{fig:v2_s1},
and Fig. \ref{fig:loss_s1} for \cnotvone{}, \cnotvtwo{}, and \cnotloss{}, respectively.

For \cnotvone{}, the plots of code performance against gate infidelity show a quadratic dependence for low $p_\text{init}$. In this case, imperfections in the \cnot{}s provide the primary source of noise. The quadratic dependence is due to the fact that at least two errors need to occur for a logical error.

The dependence becomes linear by at least ${p_\text{init}=0.002}$, with a value of $p_\text{code} \sim 10^{-4}$ at zero gate infidelity. The latter feature is due to logical errors occurring caused entirely by initialization noise.
The former is due to logical errors composed of a single initialization error and a single error from a \cnot{}.
There is no longer any hint of the quadratic dependence,
showing that logical errors due to only to pairs of \cnot{} errors have become rare in comparison.

By $p_\text{init}=0.07$, any dependence of $p_\text{code}$ on the gate infidelity has essentially disappeared.
The initialization noise is therefore the dominant source of logical errors in this case.

For each value of the infidelity, there is a range of different $p_\text{code}$ that can result.
This is most easily seen for $p_\text{init}=0$.
For low infidelity, the corresponding values of $p_\text{code}$ can differ by as much as a factor of two.
This factor decreases for larger infidelity, but still remains sizable.

This shows that, if we know only the gate infidelity for a \cnotvone{} gate,
we cannot form a highly accurate prediction of the $p_\text{code}$ it will realize,
since the specific structure of the infidelity will also play a role.
Instead, the infidelity can only provide a lower-bound estimator for the code performance.

The results for \cnotvtwo{} show similar qualitative features.
The main difference is that the range of possible $p_\text{code}$ for any given infidelity is much smaller in this case.
For this type of gate, the fidelity is therefore a better indicator of code performance.

For the coupling parameters we consider, the results for \cnotloss{} cover a range of much higher gate infidelities.
Much higher values are therefore also seen for $p_\text{code}$.
The loss of the quadratic scaling therefore occurs much earlier.
The spread of the curve is found to be quite small in this case,
and so the fidelity is found to be a good indicator of the code performance for such a \cnot{}.

\subsection{Results for Scenario II}

\begin{figure}[htbp]
    \begin{subfigure}{0.32 \columnwidth}
        \includegraphics{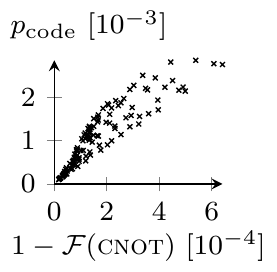}
        \includegraphics{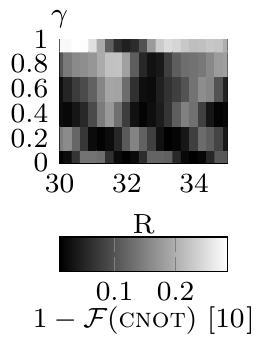}
        \caption{$p_\text{init}=0$}
    \end{subfigure}
    \begin{subfigure}{0.32 \columnwidth}
        \includegraphics{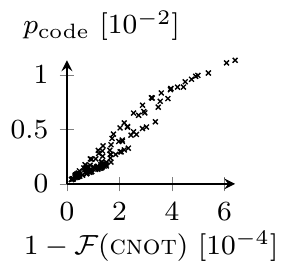}
        \includegraphics{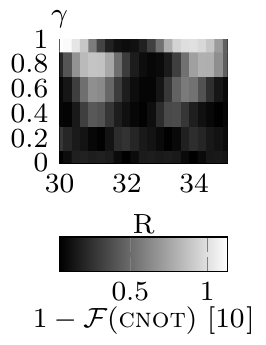}
        \caption{$p_\text{init}=0.002$}
    \end{subfigure}
    \begin{subfigure}{0.32 \columnwidth}
        \includegraphics{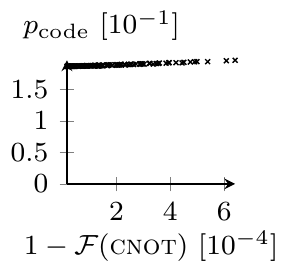}
        \includegraphics{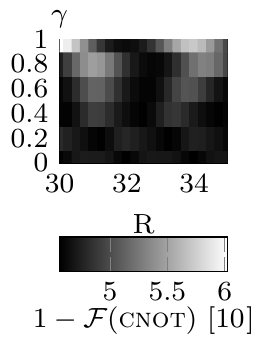}
        \caption{$p_\text{init}=0.07$}
    \end{subfigure}
    \caption{\label{fig:v1_s2}
    Plots of $p_\text{code}$ for \scenariotwo{} versus the infidelity for \cnotvone{}.
    Different curves correspond to different values of $p_\text{init}$.}
\end{figure}

\begin{figure}[htbp]
    \begin{subfigure}{0.32 \columnwidth}
        \includegraphics{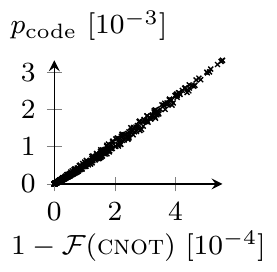}
        \includegraphics{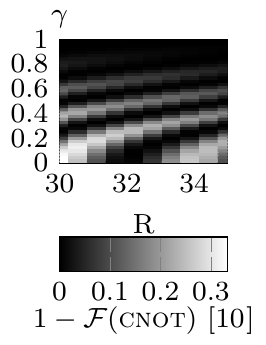}
        \caption{$p_\text{init}=0$}
    \end{subfigure}
    \begin{subfigure}{0.32 \columnwidth}
        \includegraphics{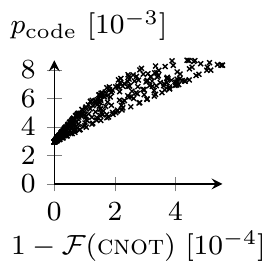}
        \includegraphics{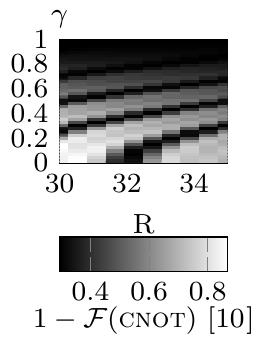}
        \caption{$p_\text{init}=0.007$}
    \end{subfigure}
    \begin{subfigure}{0.32 \columnwidth}
        \includegraphics{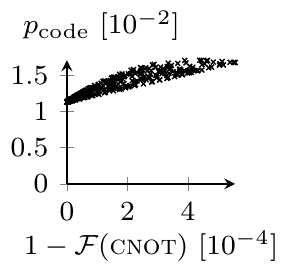}
        \includegraphics{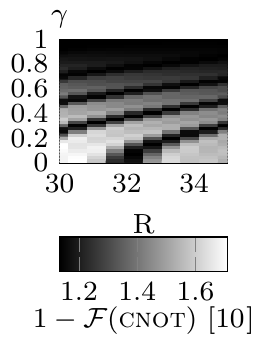}
        \caption{$p_\text{init}=0.014$}
    \end{subfigure}
    \caption{\label{fig:v2_s2}
    Plots of $p_\text{code}$ for \scenariotwo{} versus the infidelity for \cnotvtwo{}.
    Different curves correspond to different values of $p_\text{init}$.}
\end{figure}

\begin{figure}[htbp]
    \begin{subfigure}{0.32 \columnwidth}
        \includegraphics{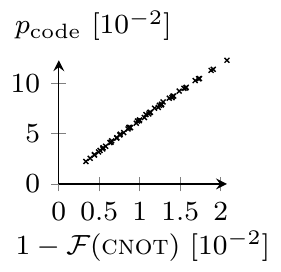}
        \includegraphics{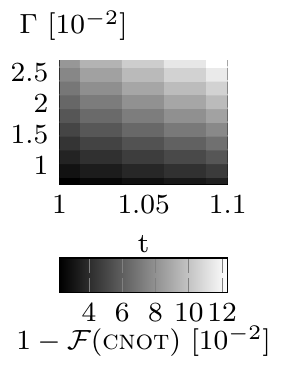}
        \caption{$p_\text{init}=0$}
    \end{subfigure}
    \begin{subfigure}{0.32 \columnwidth}
        \includegraphics{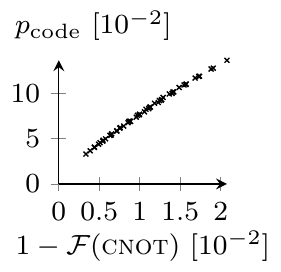}
        \includegraphics{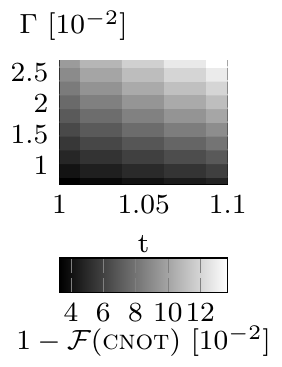}
        \caption{$p_\text{init}=0.003$}
    \end{subfigure}
    \begin{subfigure}{0.32 \columnwidth}
        \includegraphics{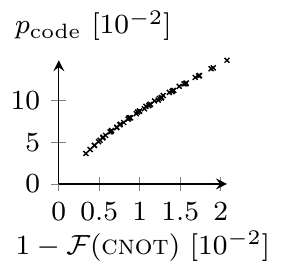}
        \includegraphics{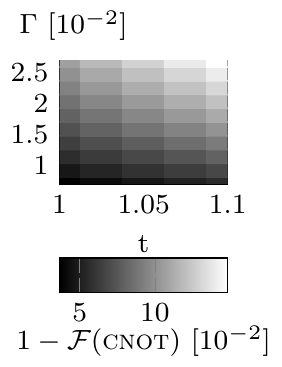}
        \caption{$p_\text{init}=0.007$}
    \end{subfigure}
    \caption{\label{fig:loss_s2}
    Plots of $p_\text{code}$ for \scenariotwo{} versus the infidelity for \cnotloss{}.
    Different curves correspond to different values of $p_\text{init}$.}
\end{figure}

We will now look at the code performance as a function of the gate infidelities for \scenariotwo{}.
The dependence of $p_\text{code}$ for this case on the infidelity of a \cnot{} is shown in
Fig. \ref{fig:v1_s2}, Fig. \ref{fig:v2_s2} and Fig. \ref{fig:loss_s2} for \cnotvone{},
\cnotvtwo{} and \cnotloss{}, respectively.

For this scenario, a logical error can already occur with only single qubit errors.
It is therefore a linear dependence of $p_\text{code}$ on gate infidelity that can be expected when this noise is dominant.

As with \scenarioone, the dependence of the code performance on the coupling parameters still generally mirrors
the dependence of the gate infidelity on the parameters, but it is not a highly accurate predictor.
In fact, we find a much greater spread of different values of code performance for a given gate fidelities for \scenariotwo{}.
 
For \cnotvone, this effect is most pronounced for low $p_\text{init}$.
As the initial noise increases, the relative importance of the
\cnot{} diminishes yet again and we recover a weak linear dependence
on the gate fidelity for high initial noise.

For \cnotvtwo{} we find that the greatest
spread in code performance is not for vanishing initial noise, as with \cnotvone. Instead it reaches a
maximum around $p_\text{init} \approx 0.007$.

Especially interesting in both cases is that the worst code
performance is not always achieved for the highest gate infidelities.
These results therefore serve as a powerful demonstration of cases for which gate fidelity serves
as a poor indicator of code performance.

The results for \cnotloss{} do not show such features.
The value of $p_\text{code}$ simply increases with the linear scaling expected for \scenariotwo, and with little spread.
The dependence on $p_\text{init}$ is minimal, since the gate infidelity is the dominant source of noise.

\subsection{Effects of the $\mathcal{K}_2$ Term}

\begin{figure}[htbp]
	\begin{subfigure}{0.49 \columnwidth}
        \includegraphics{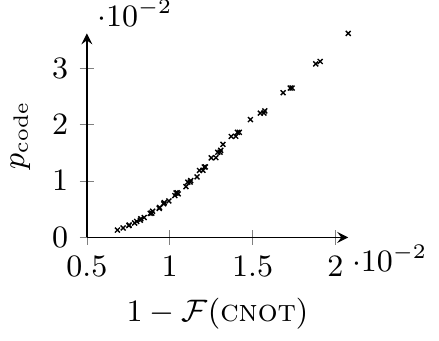}
		\caption{\cnotloss{} with $\mathcal{K}_2$ term.}
	\end{subfigure}
	\begin{subfigure}{0.49 \columnwidth}
        \includegraphics{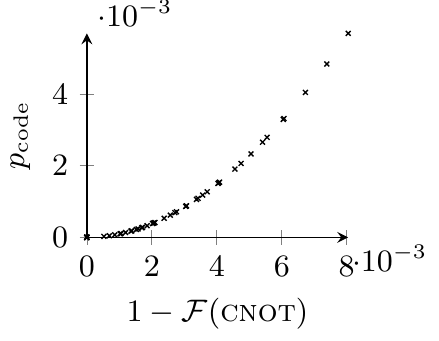}
		\caption{\cnotloss{} without $\mathcal{K}_2$ term.}
	\end{subfigure}
	\caption{$p_\text{code}$ for \scenarioone{} as a function of the all channel infidelities found within the parameter regime studied for \cnotloss{}.}
	\label{fig:loss_vs_loss-k2}
\end{figure}

For the noisy implementation of the Heisenberg coupling, such as that in \cnotloss{},
the effects of the $\mathcal{K}_2$ term are often ignored for simplicity~\cite{watson_programmable_2018}.
To assess the impact of this approximation, we consider the effects of the noise on the code both with and without this term.
The results are shown in Fig. \ref{fig:loss_vs_loss-k2}.

From these results it is evident that the $\mathcal{K}_2$ term has a significant effect.
Within the parameter regime studied, infidelities are found as high as $2 \times 10^{-2}$ with the $\mathcal{K}_2$ term,
but only as high as $8 \times 10^{-3}$ without it.
Though these values are of the same order, the difference is nevertheless significant.

Only a few values for the results with $\mathcal{K}_2$ fall within the same range as those without it.
In each case we find that, for a given fidelity, \cnotloss{} without $\mathcal{K}_2$ has a larger $p_\text{code}$ than that with.

We therefore find that neglecting the $\mathcal{K}_2$ term leads to a significant overestimate of the fidelity of
an instance of \cnotloss{}.
However, we also note that \cnot{}s of a given fidelity obtained without the $\mathcal{K}_2$
term have noise that is more damaging to the surface code than one with it.

\section{Conclusions}

In this work we proposed benchmarks for approximate \cnot{} gates.
These directly assess their capability to implement the surface code,
and do so using a system small enough to allow straightforward calculation.

The two measures of \cnot{} performance proposed were then calculated and compared to the standard
gate infidelity for a range of different implementations of the \cnot{} with spin qubits.

Our results showed that the infidelity cannot be regarded as a good predictor of code performance, beyond a simple rule of thumb.
We also found very different behaviour for both of the surface code tasks,
showing that our metrics give different perspectives on the effectiveness of a \cnot{} within the code.
This shows that, even within the limited realm of tasks involving the surface code,
the quality of a \cnot{} cannot be easily reduced to a single number.

Our benchmarks are defined by assessing the response of the surface code to specific sources of noise.
This is primarily the noise that arises during the implementations of the \cnot{}s.
Since these conditions are not easily replicable experimentally,
our benchmarks are primarily designed to be calculated using either a
theoretical description of the imperfections within a \cnot{}, or data from a quantum process tomography.
This makes our approach applicable for a wide range of theoretically described or experimentally implemented \cnot{}s,
regardless of whether they could be used in a 17-qubit surface code in the near term.

\section{Acknowledgements}

The authors would like to thank the SNSF and NCCR QSIT for support.

\bibliography{main-arxiv.bbl}

\begin{thebibliography}{14}
\expandafter\ifx\csname natexlab\endcsname\relax\def\natexlab#1{#1}\fi
\expandafter\ifx\csname bibnamefont\endcsname\relax
  \def\bibnamefont#1{#1}\fi
\expandafter\ifx\csname bibfnamefont\endcsname\relax
  \def\bibfnamefont#1{#1}\fi
\expandafter\ifx\csname citenamefont\endcsname\relax
  \def\citenamefont#1{#1}\fi
\expandafter\ifx\csname url\endcsname\relax
  \def\url#1{\texttt{#1}}\fi
\expandafter\ifx\csname urlprefix\endcsname\relax\def\urlprefix{URL }\fi
\providecommand{\bibinfo}[2]{#2}
\providecommand{\eprint}[2][]{\url{#2}}

\bibitem[{\citenamefont{Nielsen and Chuang}(2000)}]{nielsen_quantum_2000}
\bibinfo{author}{\bibfnamefont{M.}~\bibnamefont{Nielsen}} \bibnamefont{and}
  \bibinfo{author}{\bibfnamefont{I.}~\bibnamefont{Chuang}},
  \emph{\bibinfo{title}{Quantum {Computation} and {Quantum} {Information}}},
  Cambridge {Series} on {Information} and the {Natural} {Sciences}
  (\bibinfo{publisher}{Cambridge University Press}, \bibinfo{year}{2000}), ISBN
  \bibinfo{isbn}{978-0-521-63503-5}.

\bibitem[{\citenamefont{Bravyi et~al.}(2017)\citenamefont{Bravyi, Englbrecht,
  Koenig, and Peard}}]{bravyi_correcting_2017}
\bibinfo{author}{\bibfnamefont{S.}~\bibnamefont{Bravyi}},
  \bibinfo{author}{\bibfnamefont{M.}~\bibnamefont{Englbrecht}},
  \bibinfo{author}{\bibfnamefont{R.}~\bibnamefont{Koenig}}, \bibnamefont{and}
  \bibinfo{author}{\bibfnamefont{N.}~\bibnamefont{Peard}},
  \bibinfo{journal}{arXiv:1710.02270 [quant-ph]}  (\bibinfo{year}{2017}),
  \bibinfo{note}{arXiv: 1710.02270},
  \urlprefix\url{http://arxiv.org/abs/1710.02270}.

\bibitem[{\citenamefont{Tomita and Svore}(2014)}]{tomita_low-distance_2014}
\bibinfo{author}{\bibfnamefont{Y.}~\bibnamefont{Tomita}} \bibnamefont{and}
  \bibinfo{author}{\bibfnamefont{K.~M.} \bibnamefont{Svore}},
  \bibinfo{journal}{Phys. Rev. A} \textbf{\bibinfo{volume}{90}},
  \bibinfo{pages}{062320} (\bibinfo{year}{2014}),
  \urlprefix\url{https://link.aps.org/doi/10.1103/PhysRevA.90.062320}.

\bibitem[{\citenamefont{Kouwenhoven et~al.}(1997)\citenamefont{Kouwenhoven,
  Marcus, McEuen, Tarucha, Westervelt, and
  Wingreen}}]{kouwenhoven_electron_1997}
\bibinfo{author}{\bibfnamefont{L.~P.} \bibnamefont{Kouwenhoven}},
  \bibinfo{author}{\bibfnamefont{C.~M.} \bibnamefont{Marcus}},
  \bibinfo{author}{\bibfnamefont{P.~L.} \bibnamefont{McEuen}},
  \bibinfo{author}{\bibfnamefont{S.}~\bibnamefont{Tarucha}},
  \bibinfo{author}{\bibfnamefont{R.~M.} \bibnamefont{Westervelt}},
  \bibnamefont{and} \bibinfo{author}{\bibfnamefont{N.~S.}
  \bibnamefont{Wingreen}}, in \emph{\bibinfo{booktitle}{Mesoscopic electron
  transport}} (\bibinfo{publisher}{Springer}, \bibinfo{year}{1997}), pp.
  \bibinfo{pages}{105--214}.

\bibitem[{\citenamefont{Etienne and
  Paris}(1987)}]{etienne_two-dimensional_1987}
\bibinfo{author}{\bibfnamefont{B.}~\bibnamefont{Etienne}} \bibnamefont{and}
  \bibinfo{author}{\bibfnamefont{E.}~\bibnamefont{Paris}},
  \bibinfo{journal}{Journal de Physique} \textbf{\bibinfo{volume}{48}},
  \bibinfo{pages}{2049} (\bibinfo{year}{1987}).

\bibitem[{\citenamefont{Loss and DiVincenzo}(1998)}]{loss_quantum_1998}
\bibinfo{author}{\bibfnamefont{D.}~\bibnamefont{Loss}} \bibnamefont{and}
  \bibinfo{author}{\bibfnamefont{D.~P.} \bibnamefont{DiVincenzo}},
  \bibinfo{journal}{Phys. Rev. A} \textbf{\bibinfo{volume}{57}},
  \bibinfo{pages}{120} (\bibinfo{year}{1998}),
  \urlprefix\url{https://link.aps.org/doi/10.1103/PhysRevA.57.120}.

\bibitem[{\citenamefont{Kloeffel and Loss}(2013)}]{kloeffel_prospects_2013}
\bibinfo{author}{\bibfnamefont{C.}~\bibnamefont{Kloeffel}} \bibnamefont{and}
  \bibinfo{author}{\bibfnamefont{D.}~\bibnamefont{Loss}},
  \bibinfo{journal}{Annual Review of Condensed Matter Physics}
  \textbf{\bibinfo{volume}{4}}, \bibinfo{pages}{51} (\bibinfo{year}{2013}),
  \urlprefix\url{https://doi.org/10.1146/annurev-conmatphys-030212-184248}.

\bibitem[{\citenamefont{Trifunovic et~al.}(2012)\citenamefont{Trifunovic, Dial,
  Trif, Wootton, Abebe, Yacoby, and Loss}}]{trifunovic_long-distance_2012}
\bibinfo{author}{\bibfnamefont{L.}~\bibnamefont{Trifunovic}},
  \bibinfo{author}{\bibfnamefont{O.}~\bibnamefont{Dial}},
  \bibinfo{author}{\bibfnamefont{M.}~\bibnamefont{Trif}},
  \bibinfo{author}{\bibfnamefont{J.~R.} \bibnamefont{Wootton}},
  \bibinfo{author}{\bibfnamefont{R.}~\bibnamefont{Abebe}},
  \bibinfo{author}{\bibfnamefont{A.}~\bibnamefont{Yacoby}}, \bibnamefont{and}
  \bibinfo{author}{\bibfnamefont{D.}~\bibnamefont{Loss}},
  \bibinfo{journal}{Phys. Rev. X} \textbf{\bibinfo{volume}{2}},
  \bibinfo{pages}{011006} (\bibinfo{year}{2012}),
  \urlprefix\url{https://link.aps.org/doi/10.1103/PhysRevX.2.011006}.

\bibitem[{\citenamefont{Trif et~al.}(2007)\citenamefont{Trif, Golovach, and
  Loss}}]{trif_spin-spin_2007}
\bibinfo{author}{\bibfnamefont{M.}~\bibnamefont{Trif}},
  \bibinfo{author}{\bibfnamefont{V.~N.} \bibnamefont{Golovach}},
  \bibnamefont{and} \bibinfo{author}{\bibfnamefont{D.}~\bibnamefont{Loss}},
  \bibinfo{journal}{Phys. Rev. B} \textbf{\bibinfo{volume}{75}},
  \bibinfo{pages}{085307} (\bibinfo{year}{2007}),
  \urlprefix\url{https://link.aps.org/doi/10.1103/PhysRevB.75.085307}.

\bibitem[{\citenamefont{Lidar et~al.}(2013)\citenamefont{Lidar, Brun, and
  Brun}}]{lidar_quantum_2013}
\bibinfo{editor}{\bibfnamefont{D.~A.} \bibnamefont{Lidar}},
  \bibinfo{editor}{\bibfnamefont{T.~A.} \bibnamefont{Brun}}, \bibnamefont{and}
  \bibinfo{editor}{\bibfnamefont{T.}~\bibnamefont{Brun}}, eds.,
  \emph{\bibinfo{title}{Quantum {Error} {Correction}}}
  (\bibinfo{publisher}{Cambridge University Press},
  \bibinfo{address}{Cambridge}, \bibinfo{year}{2013}), ISBN
  \bibinfo{isbn}{978-1-139-03480-7},
  \urlprefix\url{http://ebooks.cambridge.org/ref/id/CBO9781139034807}.

\bibitem[{\citenamefont{Fowler}(2015)}]{fowler_minimum_2015}
\bibinfo{author}{\bibfnamefont{A.~G.} \bibnamefont{Fowler}},
  \bibinfo{journal}{Quantum Info. Comput.} \textbf{\bibinfo{volume}{15}},
  \bibinfo{pages}{145} (\bibinfo{year}{2015}), ISSN \bibinfo{issn}{1533-7146},
  \urlprefix\url{http://dl.acm.org/citation.cfm?id=2685188.2685197}.

\bibitem[{\citenamefont{Wootton et~al.}(2017)\citenamefont{Wootton, Peter,
  Winkler, and Loss}}]{wootton_proposal_2017}
\bibinfo{author}{\bibfnamefont{J.~R.} \bibnamefont{Wootton}},
  \bibinfo{author}{\bibfnamefont{A.}~\bibnamefont{Peter}},
  \bibinfo{author}{\bibfnamefont{J.~R.} \bibnamefont{Winkler}},
  \bibnamefont{and} \bibinfo{author}{\bibfnamefont{D.}~\bibnamefont{Loss}},
  \bibinfo{journal}{Phys. Rev. A} \textbf{\bibinfo{volume}{96}},
  \bibinfo{pages}{032338} (\bibinfo{year}{2017}),
  \urlprefix\url{https://link.aps.org/doi/10.1103/PhysRevA.96.032338}.

\bibitem[{\citenamefont{Darmawan and
  Poulin}(2017)}]{darmawan_tensor-network_2017}
\bibinfo{author}{\bibfnamefont{A.~S.} \bibnamefont{Darmawan}} \bibnamefont{and}
  \bibinfo{author}{\bibfnamefont{D.}~\bibnamefont{Poulin}},
  \bibinfo{journal}{Physical Review Letters} \textbf{\bibinfo{volume}{119}}
  (\bibinfo{year}{2017}), ISSN \bibinfo{issn}{0031-9007, 1079-7114},
  \urlprefix\url{http://link.aps.org/doi/10.1103/PhysRevLett.119.040502}.

\bibitem[{\citenamefont{Watson et~al.}(2018)\citenamefont{Watson, Philips,
  Kawakami, Ward, Scarlino, Veldhorst, Savage, Lagally, Friesen, Coppersmith
  et~al.}}]{watson_programmable_2018}
\bibinfo{author}{\bibfnamefont{T.~F.} \bibnamefont{Watson}},
  \bibinfo{author}{\bibfnamefont{S.~G.~J.} \bibnamefont{Philips}},
  \bibinfo{author}{\bibfnamefont{E.}~\bibnamefont{Kawakami}},
  \bibinfo{author}{\bibfnamefont{D.~R.} \bibnamefont{Ward}},
  \bibinfo{author}{\bibfnamefont{P.}~\bibnamefont{Scarlino}},
  \bibinfo{author}{\bibfnamefont{M.}~\bibnamefont{Veldhorst}},
  \bibinfo{author}{\bibfnamefont{D.~E.} \bibnamefont{Savage}},
  \bibinfo{author}{\bibfnamefont{M.~G.} \bibnamefont{Lagally}},
  \bibinfo{author}{\bibfnamefont{M.}~\bibnamefont{Friesen}},
  \bibinfo{author}{\bibfnamefont{S.~N.} \bibnamefont{Coppersmith}},
  \bibnamefont{et~al.}, \bibinfo{journal}{Nature}
  \textbf{\bibinfo{volume}{555}}, \bibinfo{pages}{633} (\bibinfo{year}{2018}),
  ISSN \bibinfo{issn}{0028-0836, 1476-4687},
  \urlprefix\url{http://www.nature.com/doifinder/10.1038/nature25766}.

\end{thebibliography}

\end{document}